\documentstyle[12pt,aaspp4]{article}

\received{....}
\accepted{....}
\journalid{337}{....}
\articleid{11}{14}

\slugcomment{{\it The Astronomical Journal}, in press}

\begin{document}
   \title{Optical Spectroscopy of the unusual galaxy J2310-43\altaffilmark{1}}

   \author{Alessandro Caccianiga\altaffilmark{2} and  
Tommaso Maccacaro}

   \affil{Osservatorio Astronomico di Brera, Via Brera 28, I-20121 Milano, Italy}
\altaffiltext{1}{Based on observations made at the European Southern Observatory, La
                 Silla (Chile).}
\altaffiltext{2}{also Dipartimento di Fisica, Universit\`a di Milano, 
                 via Celoria 16, I-20133 Milano, Italy}

\begin{abstract}
We present and discuss new spectroscopic observations of the unusual
galaxy J2310-43. The observations cover a wide 
wavelength range, from 3700\AA\  to 9800\AA\ allowing the study of both the 
regions where $H\alpha$ and the Ca II ``contrast'' are expected.
No evidence for $H\alpha$ in emission is found and we thus confirm the
absence of emission lines in the spectrum of J2310-43, ruling out the possibility
that it may host a Seyfert nucleus. The CaII break is clearly 
detected and
the value of the contrast 
(38\%$\pm$4\%) is intermediate between that of a typical elliptical galaxy 
($\approx$ 50\%) and that of a BL Lac object ($\leq 25\%$). This result
imposes limits on the intensity of a possible non-stellar continuum
and, in the light of the radio and X-ray loudness of the source, 
draws further attention to the problem of the recognition of a BL Lac object.
Objects like J2310-43 may be more common than previously recognized, and begin to
emerge in surveys of radio-emitting X-ray sources.

\end{abstract}


%
%

\section{Introduction}

In a recent paper, Tananbaum et al. (1997) have reported a detailed analysis 
of a ROSAT PSPC observation of J2310-43, a very peculiar and 
interesting galaxy firstly discovered as a luminous X-ray source ($\sim 10^{44}$ erg s$^{-1}$) in an 
{\it Einstein} IPC image (Tucker, Tananbaum \& Remillard 1995).
The IPC data showed that the X-ray source is spatially extended. This, 
combined with the fact that the galaxy is a cD in cluster (Tucker, 
Tananbaum \& Remillard 1995) supported 
the hypothesis that the X-ray emission is due to the cluster, rather 
than to the activity of the galaxy itself. However, on the basis 
of the X-ray spatial and spectral analysis of the PSPC data, 
Tananbaum et al. (1997) suggested that only 20\% 
of the total X-ray emission comes from the cluster, while the bulk 
is associated with a nuclear activity of the galaxy. 
Optical spectroscopy 
of the galaxy (Tucker, Tananbaum \& Remillard 1995), however, 
did not reveal the emission lines or the excess blue continuum 
expected if J2310-43 contains an active nucleus. 

Another interesting possibility considered by Tananbaum et al. (1997), 
is that J2310-43 is related to 
the BL Lac phenomenon. The lack of emission lines in the optical spectrum would
apparently support this view. Furthermore, the broad band energy distribution
of J2310-43, i.e. the position of the object in the 
$\alpha_{ro} / \alpha_{ox}$ plane (Tananbaum et al. 1997), falls at the edge 
of the region occupied by BL Lac 
objects (Stocke et al. 1991). 
On the other hand, the same authors 
show that the (B-V) color observed in J2310-43 is consistent 
with that of a normal elliptical galaxy and it is rather different from 
that of a typical BL Lac object. The resulting picture is rather intriguing
with contradicting or inconclusive evidences as for the presence of 
``nuclear activity". Indeed Tananbaum et al. (1997) leave open the question
whether J2310-43 belongs to the tail of the BL Lac population or to a different
class of objects and point out similarities with the optically dull galaxies 
with
strong nuclear X-ray emission discovered by Elvis et al. (1981).

The 
spectroscopic data on J2310-43, however, were limited to the 4700\AA - 6700\AA
 \ 
interval thus 
excluding two regions critical for the understanding of the nature of the 
source:
that of the Ca II break (expected at $\sim$4355\AA) and of
the $H\alpha$ line (expected at $\sim$7145\AA). In fact AGNs showing no 
$H\beta$ and [OIII] lines but exhibiting a 
broad $H\alpha$ line are known
to exist (Stocke et al. 1991; see also Halpern, Eracleous \& Forster 1997). 
For these reasons the possibility
that J2310-43 is hosting a reddened Seyfert nucleus or a BL Lac object
could not be completely ruled out. 

With the aim of further studying J2310-43 and understanding its real nature
we have therefore secured two optical spectra covering the   
wavelength range 3700\AA\ $-$ 9800\AA. 

\section{Observation and analysis}

Spectroscopy of J2310-43 was carried out with the ESO 3.6m telescope 
on 1996 December 10. Observations were made with EFOSC1 in longslit mode, 
using a 1.5 arcsec wide slit and two different grisms, b300 and r300, 
with wavelength coverage
from 3700\AA\  to 6800\AA\  and from 6200\AA\  to 9800\AA, respectively. 
The dispersions achieved with the two grisms, scaled to the
Tek512 CCD detector, were 6.3 \AA/pixel (b300) and 
7.5 \AA/pixel (r300). The exposure time was of 600 sec with the grism 
b300 and of 300 sec with the r300.

The data were reduced using the IRAF-LONGSLIT package\footnote{IRAF is 
distributed by NOAO, which is operated by AURA, Inc., 
under contract to the NSF.}. 
The wavelength solution was obtained using a He-Ar reference spectrum 
while the correction for the instrument response was based on the 
observation of a photometric standard (LTT 377). We did not make  
an absolute flux calibration, thus the flux density scale of the 
spectra is in arbitrary units.  

The calibrated spectra are presented in Figure~\ref{fig1}. 

\section{Discussion and conclusions}
\begin{figure}
\plotone{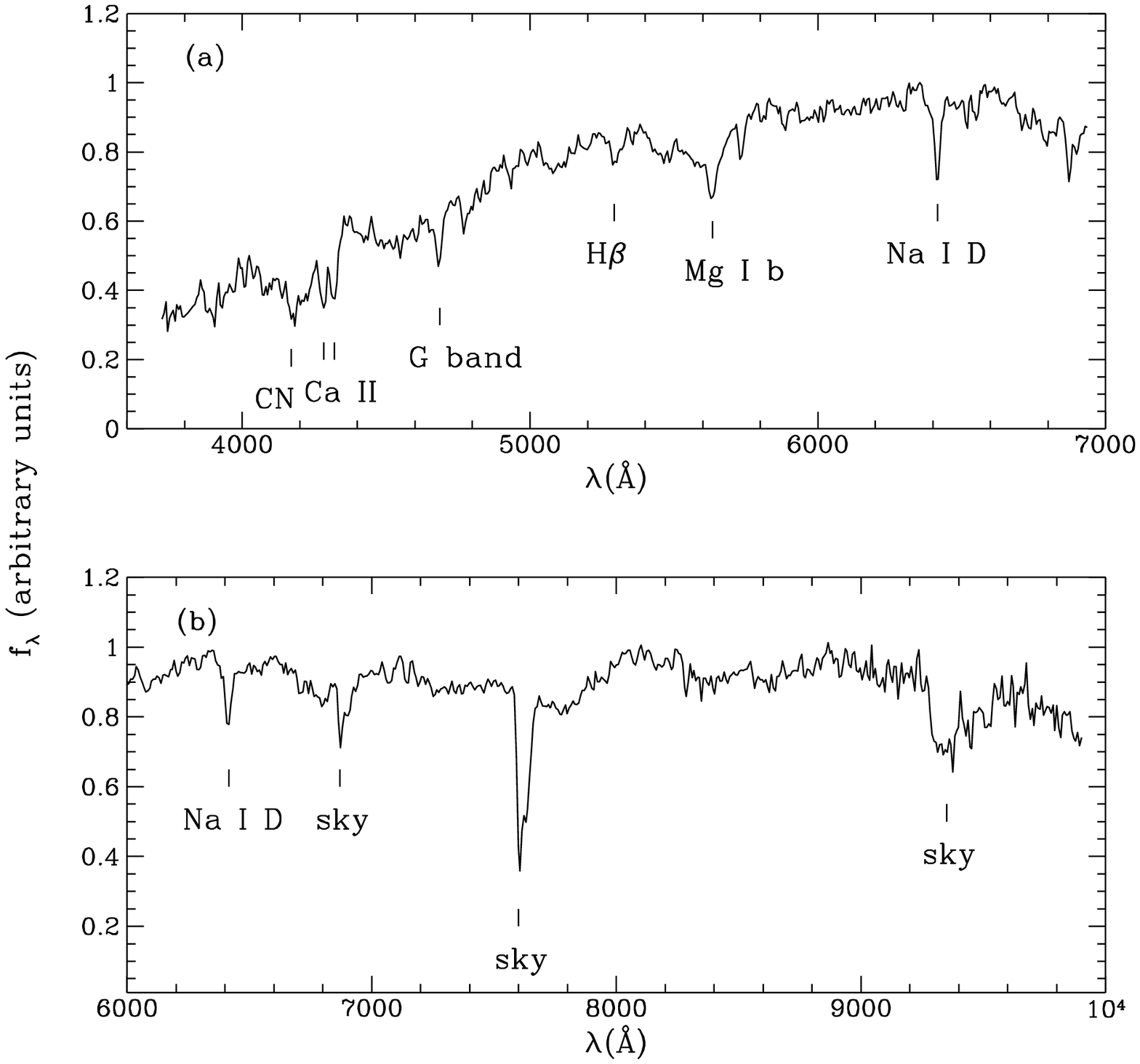}
\figcaption[caccianiga.f1.ps]{The optical spectrum of J2310-43 obtained 
with the ESO 3.6 m telescope$+$EFOSC1 with two different grisms:  b300 (a) 
and  r300 (b). The flux ($f_{\lambda}$) is in arbitrary units. \label{fig1}}
\end{figure}

The spectra presented in Figure~\ref{fig1} cover a wavelength range
considerably larger than that of the spectrum discussed in 
Tucker, Tananbaum \& Remillard (1995). In particular, the region of the 
[OII], CaII break ($\approx$ 4355\AA) and the region where H$\alpha$ is 
expected
(7145\AA) are fully covered. No emission lines appear in the spectrum, which 
shows only the typical absorption features of a ``normal'' early type galaxy.
From the main absorption features seen (Ca II H\&K, G band, H$\beta$, MgI 
5175\AA, 
Na I D) we have computed a redshift of $z=0.0887 \pm 0.0002$, 
confirming the value found by Tananbaum et al. (1997). 

The first result worth noting is the absence of H$\alpha$ in emission. 
The fact that no emission lines are present 
from [OII] to [NII] definitively put to rest the possibility that a
Seyfert nucleus is hiding in J2310-43.

Secondly, we note that a pronounced Ca II contrast is detected.  
We have computed its amplitude following 
Dressler \& Shectman (1987), i.e. by estimating the average fluxes 
(expressed in unit of frequency) between 3750\AA-3950\AA\ ($f^-$) and between 
4050\AA-4250\AA\  ($f^+$) in the rest-frame of the 
source; the contrast is then defined by:
\begin{equation}
Ca II  = \frac{f^+ - f^-}{f^+}
\end{equation}
We have found, using the central part of the spectrum to minimize the 
stellar contribution, a Ca II contrast of 38\% $\pm$ 4.0\%, which is 
below the mean value found for a ``normal'' elliptical galaxy 
($\approx$ 50\%, Dressler \& Shectman 1987).

If one considers the ``canonical'' limit of 25\% for the definition of a 
BL Lac object (Stocke et al. 1991) then J2310-43 cannot be considered a BL Lac.
On the other end March\~a et al. (1996) have proposed that 
objects with a Ca II contrast below 40\% are likely to have an extra 
source of continuum, besides stellar, and consequently they have to be 
considered as possible low-luminosity BL Lac candidates.  

The observed contrast can be used to set limits on the 
presence of such a non-stellar continuum, at least in the 
wavelength range  3700\AA \ $-$ 4300\AA \ (object rest frame). To this end, we 
have considered 
the spectrum of a ``normal'' galaxy, showing a Ca II contrast of about 60\%. 
Then, we have ``added'' a non-stellar continuum to the spectrum, 
in the form of a power-law ($f_{\nu} \propto \nu^{-\alpha}$) with a spectral 
index ranging from 0 to 2, and we have computed the Ca II contrast 
as a function of the fraction of non-stellar over 
stellar continuum. Our results show that the Ca II contrast is about  
40\% if the non-stellar contribution is approximately equal to the stellar 
continuum, (integrated between 3750\AA \ and 4250\AA).
Values of the contrast of $\leq$ 25\% (the limit used to define a 
BL Lac object) 
are obtained when the non-stellar contribution is about 3 times 
or more higher than 
the stellar continuum. Thus, in J2310-43, which has a Ca II ``contrast'' 
of $\sim 38\%$, the non-thermal component can 
still be present but at an intensity level comparable or lower than 
the fraction of the stellar continuum falling in the extraction region 
within the slit aperture.   
We have also extracted the spectrum of J2310-43 considering only 
the outer region of the galaxy, thus minimizing the contribution of the 
nucleus, and we have found that the Ca II ``contrast'' 
increases to 47\%$\pm$5\%. We consider this as a further evidence that 
J2310-43 harbors in its nucleus a weak source of non-thermal continuum 
that can be detected only if the stellar contribution falling in the 
aperture is kept at a minimum. 
These results are consistent with the observed color of J2310-43 
(determined for the whole galaxy) that indicates a negligible non-stellar 
contribution in the optical band, as discussed by Tananbaum et al. (1997). 

In conclusion, the spectroscopic observations of the galaxy J2310-43 
presented here support the  interpretation that this object represents 
the faint tail of the BL Lac population, in which the extra source of 
continuum is present but does not contribute significantly to the optical 
spectrum. 

Further observations (polarization, radio spectral index, high 
resolution X-ray spectroscopy, etc.) are obviously needed to 
 characterize the presence of such non-thermal continuum emission
in J2310-43.

It is worth noting that a  similar object (E0336-248)
 has been recently discovered by Halpern et al. (1997). 
Also in this case, the computed Ca II ``contrast'' (33\%) 
does not meet the nominal criterion of $\leq$ 25\% for 
classification as a BL Lac object. Nevertheless, these authors have 
produced convincing evidence for the presence of a non-stellar 
continuum in the spectrum of E0336-248 and, consequently, for its 
classification as a weak BL Lac object. Clearly, a re-assessment 
of the Ca II ``contrast'' criterion for the definition of a BL Lac 
is needed. 

Tananbaum et al. (1997) ask ``how common are sources such as J2310-43?"
We have reason to believe that they may be more common than currently
recognized. We have recently initiated a survey of Radio Emitting X-ray
sources (the REX Survey, Caccianiga et al. 1997a, 1997b) with the aim of 
selecting a new large sample of BL Lac objects and radio loud quasars. 
To enter the sample,
a source has to be detected in a pointed ROSAT PSPC observation {\it and}
in the VLA NVSS survey, above well defined flux limits 
and thresholds. 
During the optical identification program of the REX sources we have already 
found six
objects that are similar to J2310-43. The properties
of these objects will be presented and discussed in detail elsewhere; here we
recall that, like J2310-43, they have X-ray luminosity in the range
$10^{43}-10^{45}\ erg\ s^{-1}$ (0.5 - 2.0 keV) and  radio luminosity in the range 
$3\times 10^{30}-3\times 10^{31}\ erg\ s^{-1} Hz^{-1}$ (1.4 GHz). They are all
radio loud ($\alpha_{ro} >$ 0.35) and they all have a CaII ``contrast" 
between 25\% and 40\% and no emission lines in their spectrum. They do not
seem to lie preferably in a cluster environment.
We note that these six objects have been found
out of $\sim 100$ new spectroscopic identifications. 
It is unfortunate however that, given the very low identification rate so 
far obtained for the REX survey, we cannot, at present, make an estimate of 
the space density of these objects. 

\acknowledgments
We thank H. Tananbaum for useful comments on an earlier version of this 
paper and A. Wolter for bringing to our attention the work on 
E0336-2248. 
This work has received partial financial support from the Italian 
Space Agency (ASI contract 95-RS-72 Dmd. 161FAE/2, ASI ARS-96-71 Dmd. 92).


\newpage

\begin{thebibliography}{}

\bibitem{} Caccianiga, A., Della Ceca, R., Maccacaro, T., \& Wolter, A.,1997a, submitted to ApJ
\bibitem{} Caccianiga, A., Della Ceca, R., Maccacaro, T., Ruscica, C., \& Wolter, A. 1997b, Mem. Soc. Astr. It. vol. 68, 325
\bibitem{} Dressler, A., \& Shectman S. 1987, AJ, 94, 899
\bibitem{} Elvis, M. at al. 1981, ApJ, 246, 20
\bibitem{} Halpern, J.P., Eracleous, M., \& Forster, K. 1997, AJ in press
\bibitem{} March\~a, M.J.M., Browne, I.W.A., Impey, C.D., \& Smith, P.S. 1996,
 MNRAS 281, 425
\bibitem{} Stocke, J.T. et al. 1991, ApJS, 76, 813
\bibitem{} Tananbaum, H., Tucker, W., Prestwich, A., \& Remillard, R. 1997, 
ApJ 476, 83
\bibitem{} Tucker, W., Tananbaum, H., \& Remillard, R. 1995, ApJ, 444, 532

\end{thebibliography}
\end{document}